\documentclass[aps,a4paper,%12pt,
rpsfig]{revtex4}

\usepackage{bm}
\usepackage{mathrsfs}
\usepackage{xcolor,color,graphicx,graphics}
\usepackage[all]{xy}
\usepackage{epsfig,subfigure}
\usepackage{latexsym,amssymb,amsmath,amsfonts} %\usepackage{amstex}
\usepackage[english]{babel} %, spanish, portuguese
\usepackage[OT1]{fontenc}
\usepackage[latin1]{inputenc}
\usepackage{makeidx}
\usepackage{hyperref}
\usepackage{color,graphicx,graphics,wrapfig,epsf}%,psfig

\begin{document}
\title{Possible scenario of relic wormhole formation}

\author{A.A. Kirillov}
%\email{kirillov@bmstu.ru}
\affiliation{Bauman Moscow State Technical University, Moscow,  105005, Russian Federation}

\author{E.P. Savelova}
%\email{sep$\_$ 22.12.79@inbox.ru}
\affiliation{Bauman Moscow State Technical University, Moscow,  105005, Russian Federation}

\date{%\today
}
%%%%%%%%%%%%%%%%%%%%%%%%%%%%%%%%%%%%%%%%%

\begin{abstract}
It is known that in the presence of virtual wormholes, the vacuum is unstable, which leads to, in general, an infinite number  of phase transitions in the very early Universe. Then the standard Kibble scenario predicts the formation of defects such as domain walls. An unusual feature of virtual wormholes is that they may generate defects with negative energy density. Such defects have macroscopic dimensions and can already support the stability of the throats of real wormholes.  This gives reason to consider relic wormholes as possible realistic astrophysical objects.
\end{abstract}

\maketitle

\section{Introduction}

It is known that the possibility to create a stable wormhole requires the presence of exotic matter that violates energy conditions \cite{MThorne}-\cite{Lo}. 
The situation with the possible existence of relic wormholes is much simpler in open models of the universe. The space of a constant negative curvature was shown to admit the existence of stable wormholes without attracting any exotic matter. Such wormholes can be obtained simply by a factorization of the Lobachevsky space over a discrete subgroup of the group of motion  \cite{KS20a,KS16}. However, such wormholes have the genus $n\geq 1$ and do not possess spherical symmetry. The simplest wormhole of this kind has the shape of a torus, and when it also has a magnetic field, it allows the formation of ring-type structures in the Universe, e.g.,  see \cite{KS20b}. Such structures, for example, widely observed ring galaxies or recently discovered unexpected circular radio objects \cite{N21}-\cite{N23} can be interpreted as indirect evidence for the presence of wormholes in the form of a torus.
%%%%%%%%%%%%%%%%%%%%%%%%%

However, our universe is flat, and therefore stable relic wormholes in the form of a torus can exist only in voids\footnote{In voids the matter density is less than the critical value and, therefore, the mean curvature there is negative. } and have gigantic (of the order of a galaxy or more) dimensions. On smaller scales space is flat, and even torus-shaped wormholes require negative energy density for their stability. Without such matter, wormholes collapse into black holes. Although in flat space the collapse of torus-shaped wormholes may have a rather low velocity \cite{KS16}, the absence of exotic matter which is capable of stabilizing wormholes represents  the main reason why most of astrophysicists reject even the possibility of the existence of such objects in our Universe. Indeed, from the point of view of  experimental physics, violation of energy conditions can be easily achieved only in quantum systems at rather small scales. For example, such a  violation can be realized for intermediate states or in virtual processes. Virtual processes occur on extremely small scales and, therefore, cannot be directly manifested in the astrophysical picture. All that quantum theory can reliably predict is the possible existence of virtual wormholes on the Planck scale.

%%%%%%%%%%%%%

There exists a number of  modified theories of general relativity in which violation of energy conditions is allowed, e.g., which are inspired by the superstrings theory, M-theory, etc.. Such theories are however phenomenological in nature and are gaining significant popularity due to inconsistencies in the modern picture of the Universe. For example, due to the absence of clear idea what is the nature of the dark matter and dark energy phenomena. Nevertheless, even string theory is not accepted as a true physical model in particle physics, although it is undoubtedly of great interest from the mathematical standpoint. The same can be said about almost all modern modified theories of general relativity. In other words, the modern verdict of experimental physics is that it is impossible to violate the energy conditions necessary for the existence of stable macroscopic or astrophysical wormholes. Of course, this does not apply to virtual processes and virtual wormholes, which exist on very small Planck scales \cite{KS22,H,Loss2} see also recent review in \cite{HTS18} and references therein. Thus, we see that the main problem here is to find a reliable physical mechanism that allows us to transfer negative energy from very small Planck scales to sufficiently large macroscopic or astrophysical scales and form relic wormholes.
%%%%%%%%%%%%%%%%

\section{The role of virtual wormholes in the Pauli-Villars regularization scheme }

It turns out that such a natural mechanism really exists in the physics of wormholes. It really makes it possible to form sufficiently large macroscopic regions of space with violation of energy conditions and, therefore, gives the first rigorous scheme for the formation of primordial wormholes in the early Universe. This mechanism is based on phase transitions in the presence of virtual wormholes. The basic idea is as follows. Scattering of free particles on virtual wormholes quite naturally leads to the emergence of additional excitations, which automatically implement the Pauli-Villars regularization scheme (PVRS) \cite{KSPV}. 
We point out that the idea in which  virtual wormholes generate additional masses for particles was first pointed out  in \cite{Loss2}. Nevertheless, this idea was not developed further, e.g., see \cite{HTS18}, and the fact that additional excitations generated by wormholes lead directly to PVRS was missed.

It is remarkable that if in PVRS all additional fields and corresponding masses have an auxiliary non-physical character, and the standard PVRS is considered as a purely formal scheme, then in the presence of virtual wormholes, the new particles are realistic particles and have the direct physical meaning. They have enormous energies (starting from the Planck value), but in principle they can be observed in laboratory experiments, at least in the future. It was demonstrated, e.g., see   \cite{KSPV} and references therein, that scattering on virtual wormholes leads to an infinite series of additional excitations. Some part of such excitations demonstrates instabilities. This means that in the early Universe the development of the metric was accompanied with, in general, an infinite series of phase transitions. The situation seems to be analogous to that in spin glasses.
The main feature of PVRS is that some of the auxiliary fields carry negative energy, i.e. they have an incorrect sign in action (otherwise it would be impossible to eliminate divergences and,  in particular, to enforce the vacuum energy density to have a finite value.). Indeed, consider for illustration a massless field and only one additional auxiliary mass. In general, the Pauli-Villars propagator has the form \cite{reg,reg2}
\[
G=\frac{1}{k^{2}}\prod\limits
_{i}\frac{M_{i}^{2}}{\left(
k^{2}+M_{i}^{2}\right) }
%=\frac{1}{k^{2}}\prod\limits_{i}\left( 1-\frac{k^{2}}{\left( k^{2}+M_{i}^{2}\right) }\right)
=\frac{1}{k^{2}}-\sum_{i}\frac{1}{
\left( k^{2}+M_{i}^{2}\right) }+...,
\]%
which in the case of a single auxiliary field is simply
\begin{equation}
G=\frac{M^{2}}{k^{2}\left( k^{2}+M^{2}\right) }=\frac{1}{k^{2}}-\frac{1}{%
\left( k^{2}+M^{2}\right) }.  \label{GF}
\end{equation}%
The formal scheme assumes that the limit of $M\rightarrow\infty$ is taken into account at the end of the calculations, and the negative sign does not require any interpretation \cite{reg,reg2}. If we are dealing with realistic particles, the negative sign must be interpreted somehow.
%%%%%%%%%%%%%

In the presence of virtual wormholes, the interpretation of the sign is straightforward. Indeed, a virtual wormhole is a Euclidean configuration describing a closed daughter universe that first separates and then joins the mother Universe. The size of the daughter universe has the Planck value, which gives the Planck mass $M\sim M_{pl}$ for additional particles. In other words, additional heavy particles correspond to the same field particles inside the daughter universes, that is, excitations in the necks of virtual wormholes. We stress that due to virtual nature of daughter universes such additional particles correspond not to a particular daughter but rather to all possible daughters, that is, they represent collective phenomenon.
Consequently, when some part of the energy of the initial field is transferred to such particles, i.e. to the daughter universes, it formally disappears from the mother Universe (at least for a short period of time of the Planck order), and then a negative sign of such particles is required to preserve the total energy.
%%%%%%%%%%%

Indeed, the conservation law has the local differential form $\frac{1}{\sqrt{-g}}\partial_{\mu}\sqrt{-g}t_{\nu}^{\mu}=0$, where $t_{\nu}^{\mu }$ is the energy-momentum tensor (or pseudotensor in general relativity). Then Gauss's theorem for a multi-connected boundary bounding a solid 4-volume gives
\begin{equation}
\sum_{A}P_{\nu }^{A}=\sum_{A}\oint_{\Sigma _{A}}t_{\nu }^{\mu }d\Sigma _{\mu
}=0,
\end{equation}%
where the index $A$ lists all the fragments of the boundaries, and the normal vectors point beyond the 4-volume. Thus, the integral conservation law looks like this
\begin{equation}
\sum_{A}S_{A}P_{\nu }^{A}=0,
\end{equation}%
where $S_{A}=n_{A}^{\nu}n_{\nu}=\pm 1$ depending on the direction of the time variable $n_{\nu}$ and the corresponding normal to the closed surface $\Sigma _{A}$. In particular, if the initial section corresponding to the moment of time $t=t_0$ contains only the mother universe, while the final section contains both the mother and daughter universes, then energy conservation takes the form
\begin{equation}
E_{Mather}^{in}=E^{out}_{Mather}+E_{baby}.
\end{equation}%
We point out that in the case of an arbitrary section $t=const$, the number of virtual child universes is generally infinite, which can make a negative contribution to the energy of the mother Universe
\[
E_{Mather}(t)=E_{Mather}(t_0)+\sum_A E_{baby}^A(t_0)-\sum_{A'} E_{baby}^{A'}(t).
\]

The virtual nature of the child universes means that the conservation of energy in the mother Universe works only for sufficiently long periods of time in accordance with the uncertainty principle $\Delta E\Delta t>1$. We also recall that from the point of view of the daughter universes, such additional particles have positive energy, as it should be. Therefore, they do not assume abnormal statistics or any exotic behavior.

We point out that the origin of such negative energy corrections to the action implies not only a modification of the momentum energy tensor. The fact is that these corrections appear due to the nontrivial topology of space-time and, therefore, have a universal nature. In particular, similar corrections appear for gravitons. This means that terms with negative energy cannot be used directly in Einstein's equations, since the action of general relativity undergoes a similar modification. Roughly, we can say that additional massive particles serve, first of all, as a source of gravity inside the daughter universes. However, since each child universe describes a virtual process (it splits off and then joins the mother universe), the real situation is more complicated, and small scales do have some effect on large scales, and vice versa.

\section{Phase transitions induced by virtual wormholes}
As was demonstrated in \cite{KSPV}, corrections to the action induced by virtual wormholes  make quantum theory finite in all orders of perturbation theory, since it accurately reproduces PVRS with an infinite number of auxiliary masses. Nevertheless, formally, the theory remains non-renormalizable, for it contains an infinite number of parameters (e.g., additional masses of particles). However, direct calculations show that the situation is not so simple. The problem is that the vacuum is unstable with respect to the creation of particles in some of additional modes of the field, which leads to the phase transitions and formation of domain walls and other possible defects. Indeed, in the case of a massless scalar field, it was found that the effective Euclidean action has the form \cite{KSPV}%
\begin{equation}
\Gamma (\varphi )=\int \left[ \frac{1+Z}{2}\left( \nabla \varphi \right)
^{2}-\frac{B}{2}\left( \nabla ^{2}\varphi \right) ^{2}-\frac{C}{4}\left(
\nabla \varphi \right) ^{4}\right] d^{4}x+...,  \label{ac1}
\end{equation}%
where
\[
Z=\frac{\pi ^{2}}{2}<a^{2}X^{2}>n,\ B=\frac{\pi ^{2}}{16}<a^{2}X^{4}>n,\ \ C=%
\frac{\pi ^{4}}{36}<a^{4}X^{4}>n
\]%
are parameters that are expressed in terms of the average distribution of virtual wormholes in the vacuum state $\rho(a,X)$. Here $a$ is the radius of the mouth of the wormhole (i.e., the radius of the daughter universe), $X$ is the distance between the entrances (i.e., between the points where the child universe branches off and connects to the mother universe), and $n=\int\rho (a,X)dad^{4}X$ is the average vacuum 4-density of wormholes. We also denote $<f>=$$\frac{1}{n}\int f\rho(a,X)dad^{4}X$ -- the average vacuum values of the wormhole parameters. The last term in (\ref{ac1}) corresponds to self-interaction, which also manifests itself in action when particles are scattered by virtual wormholes. The action can be rewritten as follows
\begin{equation}
\Gamma (\varphi )=\int \left[ \frac{1+Z}{2}\varphi \left( -\square \frac{%
(-\square +\left( -M_{1}^{2}\right) )}{\left( -M_{1}^{2}\right) }\right)
\varphi -\frac{C}{4}\left( \nabla \varphi \right) ^{4}\right] d^{4}x+...,
\label{efact}
\end{equation}%
where the value of the additional mass is $M_{1}^{2}=\frac{1+Z}{B}$ and $\square =\nabla^{2}$. The constant parameter $Z$ renormalizes the value of the field $\varphi$ and can be adsorbed in it. This action directly leads to the Pauli-Villars propagator  in the form (\ref{GF}). However, we see that already the first additional mass here has an incorrect sign $M^{2}=\left( -M_{1}^{2}\right) <0$, which reflects the instability of the vacuum state $<0|\varphi|0>=<0|\nabla \varphi |0>=0$ with respect to the birth of scalar particles propagating through the virtual daughter universes. The instability of the vacuum leads to a series of phase transitions in the very early universe (in general, phase transitions occur for each unstable mode). During this phase, a large number of particles are born which leads to a specific polarization of vacuum which can be described by a classical field. We point out that in general, since the number of unstable modes is infinite, we have to introduce an infinite number of classical fields (the order parameters) which will describe the structure of the stable vacuum state.  This corresponds to the non-renormalizability of quantum gravity.

To consider only additional degrees of freedom (e.g., particles propagating through virtual babies), we can introduce an auxiliary potential vector field as follows $\nabla\varphi=\frac{M_{1}}{\sqrt{1+Z}}\mathbf{u}$. Then the action (\ref{efact}) becomes
\begin{equation}
\Gamma (\varphi )=-\int \left[ \frac{1}{2}\mathbf{u}\left( -\square \right)
\mathbf{u}-\frac{1}{2}M_{1}^{2}\mathbf{u}^{2}+\frac{\lambda }{4}\mathbf{u}%
^{4}\right] d^{4}x+...,  \label{efact2}
\end{equation}%
where $\lambda =C\left( \frac{M_{1}}{\sqrt{1+Z}}\right) ^{4}$. In terms of the $\mathbf{u}$ field, the action has a negative sign that reflects the negative energy of the corresponding particles from the point of view of the mother universe.  Potential energy for the auxiliary field $\mathbf{u}$ provides a mechanism for spontaneous symmetry breaking and has the form
\[
U(\mathbf{u})=-\frac{M^{2}}{2}\mathbf{u}^{2}+\frac{\lambda }{4}\mathbf{u}%
^{4}=\frac{\lambda }{4}\left( \mathbf{u}^{2}-u_{0}^{2}\right) ^{2}-\frac{%
\lambda }{4}u_{0}^{4},
\]%
where $u_{0}^{2}=M_{1}^{2}/\lambda$. Here the last term gives a positive contribution to the cosmological constant $\Lambda =\frac{\lambda}{4}u_{0}^{4}$. Note that the cosmological constant contributes only to the evolution of the scale factor of the Universe.
The minimum of potential energy defines a new stable vacuum state, which corresponds to a non-zero value $<0|\mathbf{u}|0>=u_{0}\mathbf{m}$, where $\mathbf{m}$ is an arbitrary unit vector, while the new mass term becomes anisotropic, but already has the correct behavior
\[
M_{ij}^{2}=\frac{\partial ^{2}U}{\partial u^{i}\partial u^{j}}%
=M^{2}m_{i}m_{j},
\]%
where $M^{2}=2M_{1}^{2}$, and the matrix $M_{ij}^{2}$ is positive definite. Since here the vector field $\mathbf{u}$ has a potential character, then the field $\mathbf{u}$  has, in fact, only one longitudinal component. It is easy to see that such a vacuum state violates the Lorentz symmetry for additional particles. Recall that, in fact, these particles correspond to collective field excitations in daughter closed universes, and, consequently, the violation of Lorentz symmetry is quite natural for them.
%%%%%%%%%%

Due to the potentiality of the field, it is always possible to choose a local coordinate system in which the vector $\mathbf{u}$ has one component $\mathbf{u}=\left(0,0,0,u_{z}\right) $. Then the two true values of the vacuum of the field are equal to $\mathbf{u}_{0}=\left(0,0,0,\pm\sqrt{1/\lambda}M_{1}\right) $, while the Goldstone boson is apparently missing here. There are topological defects such as domain walls that separate regions with different vacuums (i.e. with different signs $u_{0}=\pm \sqrt{1/\lambda}M_{1}$). The simplest domain wall is given by the formula
\begin{equation}
u_{z}=u_{0}th\left( \sqrt{\frac{\lambda }{2}}u_{0}z\right) ,\
u_{t}=u_{x}=u_{y}=0, \label{u}
\end{equation}%
what corresponds to the solution of the system (see details in \cite{Bog76})%
\[
\frac{\partial u_{z}}{\partial t}=\frac{\partial u_{z}}{\partial x}=\frac{%
\partial u_{z}}{\partial y}=0,\frac{\partial u_{z}}{\partial z}=\sqrt{\frac{%
\lambda }{2}}\left( u_{0}^{2}-u_{z}^{2}\right) .
\]%
The presence of infinite walls should probably be in a conflict with observations. Indeed, regions with negative energy density will expand rapidly, leading to the birth of particles already in the mother universe until the total energy density takes a positive value. However, such solutions can be used for an inflationary scenario. A more realistic situation corresponds to the domain walls located in the mouths of wormholes that connect regions with different vacuums. Here, the negative energy density is compensated by the negative curvature of the neck, which leads to a stable configuration. Consider the simplest Ellis-Bronnikov \cite{Bron73,Ellis73} (or Morris-Thorne \cite{MThorne}) metric
\[
ds^{2}=-dt^{2}+dr^{2}+\left( r^{2}+b^{2}\right) d\Omega ^{2}.
\]%
Here the radial coordinate runs through the values of $r\in (-\infty ,+\infty )$, and the minimum radius of the neck $b$ can have an arbitrary value. The domain wall type solution connects two vacuum states $u_{r}\left(\pm\infty\right) =\pm u_{0}$ and has the same shape (\ref{u}) with an obvious replacement $z\rightarrow r$. It is important that since $u_{r}\sim\partial_{r}\phi$, both vacuum states $\pm u_{0}$ correspond to the same vacuum state. Indeed, if we introduce a new coordinate $r^{\prime}=-r$ in the asymptotic domain $r\rightarrow -\infty$, this will change $u_{0}^{\prime}=-u_{0}$ and both asymptotic domains $r\rightarrow\pm\infty$ will correspond to the same vacuum state $u_{r}|_{\pm\infty}=u_{0}$. Therefore, a wormhole can connect regions with the same vacuum in the same space.

%%%%%%%%%%
For the given metric, the domain wall (\ref{u}) is located in the middle of the neck $r=0$, it has a thickness of the order of $\sim 1/M$, and the energy density of the wall is
\[
\rho _{0}=-\sqrt{\frac{\lambda }{2}}\left( u_{0}^{2}-u_{r}^{2}\right) \frac{%
\partial u_{r}}{\partial r}
\]%%%%%%%%%%%%%%%%%%
and has a negative sign. Negative energy density protects the wormhole from collapse, and the size of the neck, in principle, can have an arbitrary value. We are not discussing here a self-consistent solution of the Einstein equations in the presence of such a domain wall on the neck. The stable configuration and the corresponding dimensions of the wormhole depend on the amount of matter surrounding the neck. There is quite an extensive literature on possible solutions, for example, see \cite{Lo,Wh-sols} and the references in it. There are also more complex topological defects (such as a domain wall) that carry a smaller or even missing amount of negative energy density. Moreover, when taking into account the rest additional field modes and other types of fields, the situation becomes even more complicated. In particular, the actual number of classical fields which we have to introduce for describing all phase transitions should be infinite.
However, all these issues require further study.
%%%%%%%%%%

We expect that a certain number of primordial wormholes formed during the phase transition when the universe cooled till a critical temperature $T_{c}=2u_{0}$ in accordance with the standard Kibble scenario \cite{Kibble}. The exact value of $T_{c}$ depends on the typical parameters of the distribution of virtual wormholes in the vacuum state. Naive estimates give Planck values for typical spacetime foam parameters. Assuming $Z\gg 1$ and estimates of the type $<a^{2}X^{2}>\sim\overline{a}^{2}\overline{X}^{2}$, where  $\overline{a}$ denotes a typical value, we find
\[
T_{c}=2\sqrt{\frac{B\left( 1+Z\right) }{C}}\sim \overline{X}\sqrt{n}.
\]%
We see that the critical temperature is determined by the typical distance between the entrances into a virtual wormhole and the average 4-density of the wormholes. Recall that the quasi-classical consideration gives only an estimate of the size of the daughter universe $\overline{a}\sim\ell_{Pl}$, for example, see \cite{H} (the Euclidean action for one virtual hole is $S_{E}=3a^{2}/2\pi$), while the typical value of $\overline{X}$ can only be obtained from laboratory measurements of the possible value of the additional mass of $M$.
\[
M^{2}=2M_{1}^{2}=2\frac{1+Z}{B}\sim \frac{16}{\overline{X}^{2}}.
\]

As follows from lattice models of quantum gravity, the distribution of virtual wormholes in the space-time foam has fractal properties \cite {frac,frac1,frac2}.
This means that the values of $\overline{X}$ and $n$ may differ significantly from the Planck values. \cite{KSV21}.
Then we can consider the opposite case of $Z\ll 1$ and find
\[
T_{c}\sim \frac{1}{\overline{a}}\sim T_{Pl},
\]%
while the mass value becomes
\[
M^{2}\sim \frac{1}{Z}\frac{16}{\overline{X}^{2}}.
\]%
This really means that the typical value of $\overline{X}$ can actually be $\overline{X}\gg\ell_{Pl}$, even if the mass has the Planck order of $M\sim 1/\ell_{Pl}$.
%%%%%%%%%%

\section{Conclusions}

Thus, we see that the action for the auxiliary field $u$ has the negative sign (\ref{efact2}), which reflects the anomalous contribution to various vacuum averages from auxiliary fields in the invariant Pauli-Villars renormalization scheme. The interpretation of negative energy modes is clear. Excitations of auxiliary field particles with mass $\widetilde{M}_{1}$ correspond to particles that propagate through throats of virtual wormholes, i.e., which are actually excited in daughter universes. Such particles do not belong to a particular virtual daughter universe, but rather they represents a collective phenomenon. At each moment of time, the spatial section $t=const$ looks like a set of unrelated spaces, a mother universe and a set of baby universes (wormhole necks). The total energy is conserved, and therefore, when part of the energy enters the daughter universes, the amount of energy in the mother Universe decreases by the same amount. In other words, from the point of view of the mother Universe, the energy of excitations at the necks of virtual wormholes has a negative sign. We point out that the sign of the energy of such particles has no a direct relation to the stability of the vacuum state. Preliminary investigations \cite{KSPV} indicate that apparently all additional modes (e.g., auxiliary fields in PVRS) were initially unstable and went through phase transitions in the early Universe.  
For the formation of a stable wormhole configuration the negative energy modes cannot be used directly in Einstein's equations, since such negative energy modes manifest themselves on Planck scales, where gravitons undergo a similar modification. However, during phase transitions, the negative energy modes can form defects such as domain walls that already have macroscopic dimensions, which allows their use as stabilizers of actual relic wormholes.

%\bigskip


\begin{thebibliography}{99}

\bibitem{MThorne} M. S. Morris and K. S. Thorne,  Wormholes
in space time and their use for interstellar travel: A tool for teaching
general relativity, \emph{Am. J. Phys.} \textbf{56}, 395
(1988).

\bibitem{MTY} M. S. Morris, K. S. Thorne and U. Yurtsever,
Wormholes, Time Machines, and the Weak Energy Condition,
\emph{Phys. Rev. Lett.} \textbf{61}, 1446 (1988).

\bibitem{VisX} M. Visser,  Lorentzian wormholes: From
Einstein to Hawking, Springer-Verlag, New-York, Inc.
(1996).

\bibitem{Lo} F. S. N. Lobo,  Wormholes, Warp Drives and
Energy Conditions, Fundam. Theor. Phys. 189, pp. (2017),
Springer Nature Switzerland AG.

\bibitem{KS20a} A.A. Kirillov, E.P. Savelova, %Wormhole as a possible
%accelerator of high-energy cosmic-ray particles.
\emph{Eur. Phys. J. C} \textbf{80}, 45 (2020).



\bibitem{KS16} A.A. Kirillov, E.P. Savelova, \emph{Int. J. Mod. Phys. D}
\textbf{25}, 1650075 (2016).

\bibitem{KS20b} A.A. Kirillov, E.P. Savelova, \emph{Eur. Phys. J. C} \textbf{%
80}, 810 (2020).

\bibitem{N21} R.P. Norris, et al, \emph{Publications of the Astron. Soc. of
Australia}, \textbf{38}, E003 (2021).

\bibitem{N22} R.P. Norris, et al, \emph{Mon. Not. RAS}, \textbf{513},
1300-1316 (2022).

\bibitem{N22b} R.P. Norris, E. Crawford, P.Macgregor, \emph{Galaxies},\
\textbf{9} 83 (2021).

\bibitem{N23} K. Dolag, L.M. Boss, B.S. Koribalski, U.P. Steinwandel, M.
Valentini, \emph{The Astrophysical Journal, }\textbf{945}, 74 (2023).


\bibitem{KS22} A.A. Kirillov, E.P. Savelova, \emph{Universe}, \textbf{\ 8},
428 (2022).


\bibitem{H} S. W. Hawking, \emph{Phys. Lett. }\textbf{B 195}, 337 (1987).
S. W. Hawking, in: Quantum Gravity. In: Markov, M.A., Berezin, V.A. Frolov, V.P. (eds.) Proceedings of the 4th International Seminar, Moscow, 1987 (World Scientific, Singapore, 1988).

\bibitem{Loss2} Lavrelashvili G.V. ,  Rubakov V.A., and  Tinyakov P.G., 
%Disruption of quantum coherence upon a change in spatial topology in quantum gravity.
{\em JETP Lett.}, {\bf  46},  134 (1987); in: Quantum Gravity. In: Markov, M.A., Berezin, V.A. Frolov, V.P. (eds.) Proceedings of the 4th International Seminar, Moscow, 1987 (World Scientific, Singapore, 1988).

\bibitem{HTS18} A. Hebecker, M. Thomas and P. Soler, 
 %Euclidean Wormholes, Baby  Universes, and Their Impact on Particle Physics and Cosmology. 
{\em Front. Astron. Space Sci.} {\bf 5} : 35 (2018). 
%doi: 10.3389/fspas.2018.00035



\bibitem{KSPV} A.A. Kirillov, E.P. Savelova,
%Effective action for a free scalar field in the presence of spacetime foam,
\emph{Gen Relativ Gravit}
\textbf{47}:97 (2015).

\bibitem{reg} R.P. Feynman, Phys. Rev., \textbf{74}, (1948) 1430

\bibitem{reg2}   Pauli W.,
 Villars F., On the Invariant Regularization in Relativistic Quantum Theory. {\em Rev. Mod. Phys.}   {\bf 1949}, 21, 434.



\bibitem{Bog76} E.B. Bogomolnyi, \emph{Sov. J. Nucl. Phys.}, \textbf{24}, 449
(1976).

\bibitem{Bron73} K.A. Bronnikov, \emph{Acta Phys. Polon. A} {\bf 4}, 251 (1973).

 \bibitem{Ellis73} H.G. Ellis, \emph{J. Math. Phys.} {\bf 14}, 104 (1973).

\bibitem{Wh-sols} K.A. Bronnikov, S.V. Sushkov, \emph{Universe}, \textbf{\ 9}%
, 81 (2023).

\bibitem{Kibble} T.W. Kibble, \emph{Phys. Rep.} \textbf{67}, 183 (1980).

\bibitem{frac} V. Knizhnik, A. Polyakov, A. Zamolodchikov, \emph{Mod. Phys.
Lett. A,} \textbf{3,} 819 (1988).

\bibitem{frac1} Ambjorn, J.; Jurkiewicz, J.; Loll, R.
%The spectral dimension of the universe is scale dependent.
\emph{Phys. Rev. Lett.}, \textbf{95},
171301 ( 2005).

\bibitem{frac2} Ambjorn, J.; Drogosz, Z.; Gizbert-Studnicki, J.; Gorlich,
A.; Jurkiewicz, J.; Nemeth, D.
%Cosmic voids and filaments from quantum gravity.
\emph{Eur. Phys. J. C,} \textbf{81}, 708 (2021).



\bibitem{KSV21} A.A. Kirillov, E.P. Savelova, P.O. Vladykina, \emph{Universe}
, \textbf{7}, 178 (2021).


\end{thebibliography}
\end{document}